\documentclass[a4paper,fleqn,usenatbib]{mnras}

\usepackage{newtxtext,newtxmath}

\usepackage[T1]{fontenc}
\usepackage{ae,aecompl}

\usepackage{graphicx}   
\usepackage{amsmath}    
\usepackage{amssymb}    
\usepackage{url}		

\title{Does the second law hold at cosmic scales?}

\author[]{Manuel Gonzalez-Espinoza,$^{1}$\thanks{E-mail: manuel.gonzalez@pucv.cl}
Diego Pav\'{o}n$^{2}$\thanks{E-mail: diego.pavon@uab.es}
\\
$^{1}$Instituto de F\'{\i}sica, Pontificia Universidad Cat\'{o}lica de Valpara\'{\i}so, Casilla 4950, Valpara\'{\i}so, Chile.\\
$^{2}$Departamento de F\'{\i}sica, Universidad Aut\'{o}noma de Barcelona, 08193 Bellaterra, Barcelona, Spain.
}

\date{Accepted XXX. Received YYY; in original form ZZZ}

\pubyear{2018}

\begin{document}
\label{firstpage}
\pagerange{\pageref{firstpage}--\pageref{lastpage}}
\maketitle

\begin{abstract}
\noindent The second law of thermodynamics is known to hold at
small scales also when gravity plays a leading role, as in the
case of  black holes and self-gravitating radiation spheres. It
has been suggested that it should  as well at large scales. Here,
by a purely kinematic analysis \textemdash based on the history of
the Hubble factor and independent of any cosmological model
\textemdash, we explore if this law is fulfilled in the case of
homogeneous and isotropic universes regardless of the sign of the
spatial curvature.
\end{abstract}




\section{Introduction}
\noindent At first sight, gravity and thermodynamics appear as
practically disjoint, if not altogether unrelated, branches of
Physics. However, as a closer look reveals, this is misleading.
Recall, for instance, Tolman's law of thermodynamic equilibrium in
a stationary gravitational field \citep{Tolman,Tolman2}, Unruh's
effect \citep{Unruh-1976}, the entropy associated to event
\citep{hawking-1975,hawking-1976, Wald-1994, gibbons-hawking-1977}
and apparent  horizons, \citep{Bak-Rey-2000, Cai-Cao-2007}, the
generalized second law of thermodynamics for black holes
\citep{Bekenstein-1974, Bekenstein-1975}, the equilibrium of
self-gravitating radiation spheres \citep{sorkin1981,
diego-peter}, and, finally, the realization that the field
equations of Einstein gravity can be understood as thermodynamic
equations of state \citep{jacobson-1995,paddy-2005}.
\\  \

\noindent Therefore, the question arises: is the Universe a
thermodynamic system? To put it another way, does it follow the laws of
thermodynamics? In general relativity, and in most theories of
gravity, the conservation of energy, \textemdash i.e., the first
law \textemdash  emerges directly from the field equations of the
theory in question, but to experimentally verify whether or not it
is fulfilled at large scales appears far beyond present human
capability. Here, we do not concern ourselves with this law; we
simply assume its validity. As for the third law, we think it does
not apply at large scales since the concept of ``temperature of
the universe" has no clear meaning. It is the second law that
we are interested in.
\\   \

\noindent At this point we feel expedient to recall it. As we
witness on daily basis, macroscopic systems tend spontaneously to
thermodynamic equilibrium. This is at the very basis of the second
law. The latter encapsulates this by asserting that the entropy,
$\, S$, of isolated systems never decreases, $S' \geq 0$ and that,
at least in the last stage of approaching equilibrium, it is
concave, $ S'' < 0$ \citep{callen}. The prime
means derivative with respect to the relevant thermodynamic
variable. In this paper we shall tentatively apply this to cosmic
expansion.
\\   \

\noindent  Before going any further, we wish to emphasize that
sometimes the second law is found formulated by stating just the
above condition on $\, S'$ but not on $\, S''$. While this
mutilated version works well for many practical purposes, it is
insufficient in general. Otherwise one would observe systems whose
entropy increased without bound, which is at odds with daily
experience. In this connection, it should be noted that this can
arise in Newtonian gravity. A case in point is the gravothermal
catastrophe: The entropy of a number of gravitating point masses
confined to the interior of  a rigid sphere, of perfectly
reflecting walls, whose radius exceeds some critical value,
diverges \citep{antonov1962, lynden1968}. This uncomfortable
outcome hints that thermodynamics and Newtonian gravity are not
fully consistent with each other. Nevertheless, it can be evaded
by resorting to general relativity: The formation of a black hole
at the center of the sphere renders the entropy of the system to
stay bounded.
\\  \

\noindent Today, it is widely agreed that the entropy of the
universe (we mean that part of the universe in causal contact with
us) is overwhelmingly contributed by the entropy of the cosmic
horizon (about $\, 10^{122}$ times the Boltzmann constant).
Supermassive black holes and the cosmic microwave radiation,
alongside the cosmic sea of neutrinos, come behind by $\, 18$ and
$\, 33$ orders of magnitude, respectively. All the other sources
of entropy contribute much less \citep{egan2010}. Therefore, to
ascertain whether the universe fulfills the second law it suffices
to see whether the entropy of the cosmic horizon, $S_{h}$, comply
with $ \, S'_{h} \geq 0$ and $S''_{h} \leq 0$. In homogeneous and
isotropic universes one can define different causal horizons. We
are interested in the apparent horizon (the boundary hyper-surface
of the spacetime anti-trapped region \cite{Bak-Rey-2000}), since
it always exists in non-static universes, independently of whether
it accelerates when $t \rightarrow \infty$ or not. Neglecting
possible (small) quantum corrections, the entropy of the  horizon
is found to be proportional to the area of the latter
\citep{Bak-Rey-2000, Cai-Cao-2007}
\begin{equation}
S_{h} = k_{B} \frac{{\cal A}}{4 \, \ell^{2}_{\rm p}} \, , \label{sh1}
\end{equation}
where ${\cal A} = 4 \, \pi \tilde{r}^{2}_{{\cal A}}\, $,
$\tilde{r}_{{\cal A}}= 1/\sqrt{H^{2} \, + \, k \, a^{-2}}\,$ the
radius of the apparent horizon, $k = +1, 0, -1$ the normalized
spatial curvature parameter, and $\ell_{\rm p}$ the Planck's
length. Thereby, it follows that the universe will comply with the
second law provided that  $\, {\cal A}' \geq 0$ and $\, {\cal A}''
\leq 0$ (the prime means derivative with respect the scale factor
of the Robertson-Walker metric).
\\  \

 \noindent The aim of this paper is to explore whether this is to
be expected in view of the current data of the history of the
Hubble factor. As it turns out, our study \textemdash which is
purely kinematic \textemdash suggests that this may well be the
case and, i.e., that the universe seems to tend to a state of
maximum entropy in the long run. This outcome was suggested
earlier \citep{ferreira-pavon2016}. Here we strengthen this
suggestion by employing an ampler dataset and using a somewhat
different analysis.
\\   \

\noindent This result is also achievable from the "holographic equipartition principle",
that the rate at which the three-spatial volume of the flat universe increases with 
expansionn is proportional to the difference of the number of degrees of freedom between 
the horizon and the bulk \citep{krishna2017holographic}. However, our approach is more 
economical as we neither use temperatures at all nor assume equality beween them.
\\   \

\noindent The paper is organized as follows. Next section presents
the observational data and the best fit to them. Section III
introduces three simple parametrizations of the Hubble function in
terms of the scale factor. As we will see, the derivatives of the
area of the apparent horizon associated to these parametrizations
fulfill the inequalities expressed in the preceding paragraph.
Section IV studies, for the sake of comparison, the evolution of
the apparent horizon for a handful of cosmological models that are
known to fit reasonably well the data. Last section presents our
conclusions  and final comments.

\section{$H(z)$ data and Gaussian Process}
\noindent We shall  use the set of $\, 39 \, $ data of the Hubble
factor, $H(z)$, alongside their $\, 1 \sigma \, $ confidence
interval, in terms of the redshift, compiled by \cite{farook2017} and \cite{ryan2018},
listed in table \ref{table:H(z)data}. These data come from
different sources, also listed there, and were obtained using
differential ages of red luminous galaxies, galaxy clustering  and
baryon acoustic oscillation techniques \textemdash see references
in the said table for details. Our study does  not rely on any
cosmological model at all. However, in section IV, we compare the
predictions of some successful models about the evolution of the
area of the horizon with the outcome of our kinematic analysis.
\\  \

\noindent Regrettably, a few of $H(z)$ data are affected by
uncomfortably big $\, 1\sigma\, $ confidence intervals whereby
some statistically-based smoothing process must be applied to the
whole dataset if one wishes to draw sensible conclusions. 
We resort to the machine learning model Gaussian process, which infers a function from labelled training
data \citep{rasmussen}. This process is capable of capturing a wide range of behaviours with
only a set of parameters and admit a Bayesian interpretation \citep{zhao2018quantum}. In this study we are implementing the Gaussian process using the Wolfram Language (which includes a wide range machine learning capabilities) to be more specific we are using the Predict function with the Performance Goal based on Quality. All the numeric analysis is only based on the library given by the Wolfram Mathematica 10.4.
See \cite{seikel2012} for a deep numerical analysis, with useful
references, to this technique in a cosmological context.
\\  \

\begin{figure}
    \centering
    \includegraphics[width=0.47\textwidth]{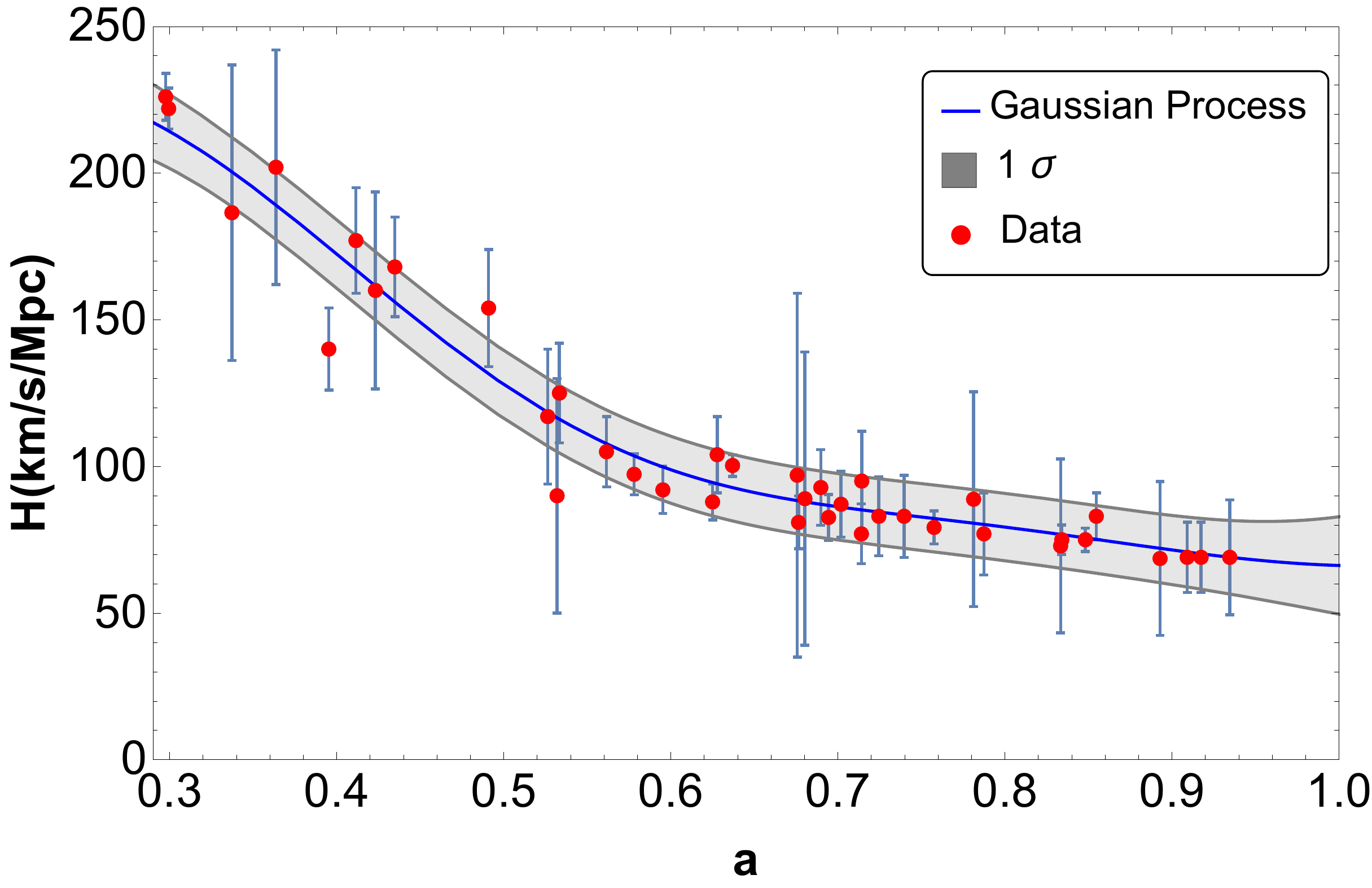}
    \caption{$H(a)$ data (red points) and their $1\sigma$ confidence interval listed in
    table \ref{table:H(z)data}. The blue solid line is the best fit
    to the data obtained by the Gaussian Process. The gray shaded band
    corresponds to the $1\sigma$ confidence interval.}
    \label{fig:gaussiang1}
\end{figure}

\noindent Application of  the Gaussian process to the dataset
results in the blue solid line \textemdash  the "best fit"
\textemdash and its $1 \sigma$ (gray shaded) confidence band, in
terms of the scale factor, as shown in Fig. \ref{fig:gaussiang1}.
Extrapolation to $\, a = 1$ gives \, $H_{0} = 66.2 \pm 16.6$
km/s/Mpc. This line suggests that $\, H'$ is negative in the whole
interval $\, 0.3 \leq a \leq 1.0\, $ and that $\, H''$ is positive
for $\, a \geq 0.4$. Inspection of Figs. 1(c) and 3(a) of \cite{carvalho2001} also supports this view. In
principle there is no grounds to believe that this should be
different for $\, a > 1$.
\\  \

\noindent Notice that $ H' < 0$ implies ${\cal A'} > 0$ (at least, in spatially flat universes). 
Likewise, $H'' > 0$  leads to  ${\cal A''} < 0$, which must be realized from 
some scale factor onwards if the entropy of the horizon is to approach a maximum in 
the long run (alongside ${\cal A'} > 0$). As we will see below, in this regard, the 
impact of the spatial curvature will have little consequence (if at all) at late times.

\section{Parameterizing the Hubble function}
\noindent Here we essay three simple parametrizations of $\, H(a)$
that comply with the  conditions $H'(a) <0$ and $H''(a) >0$,
mentioned above, in the interval covered by the $H(z)$  data
listed  in table \ref{table:H(z)data}, and that lead to ${\cal A}'
>0$ and ${\cal A}'' \leq 0$ (the latter inequality is not
necessarily valid when $a < 0.5$). We emphasize that these three
Hubble functions are not meant to describe cosmic expansion for
$\, a \ll 1$, but we hope they will be qualitatively correct
otherwise. We will contrast them with the best fit (blue solid
line) shown in Fig. \ref{fig:gaussiang1}. We will take as the
``goodness" of a given parametrization  the area enclosed by the
said best fit line and the graph of the parametrization (the
smaller the area, the better the parametrization).

\subsection{Parametrization 1}
\noindent Our first proposal is
\begin{equation}
H(a)=H_{*} \, e^{\lambda / a}\, .
\label{eq:Hmodel1}
\end{equation}
By numerically fitting it  to the best fit line with the $1\sigma$
confidence interval of Fig. \ref{fig:gaussiang1}, one obtains for
the free parameters: $H_{*}=41.15 \pm 2.30$ km/s/Mpc and $\lambda=
0.51 \pm 0.03$. Then, $H_{0}=68.5$ km/s/Mpc, and the age of the
universe, defined as $t_{0} = \int_{0}^{1}{da/(a\, H(a))}$, is $\,
13$ Gyr. On the other hand, the deceleration parameter, $q = -1 -
(aH'/H)$,  evaluated at  present gives $q_0=-0.49$ and the
transition deceleration-acceleration occurs at $a_{tr} = 0.51$.
The area between the graph of  the parametrization and the blue
solid line (best fit)   obtained using the Gaussian Process is $\,
5.15$.
\\  \

\begin{figure}
    \centering
    \includegraphics[width=0.45\textwidth]{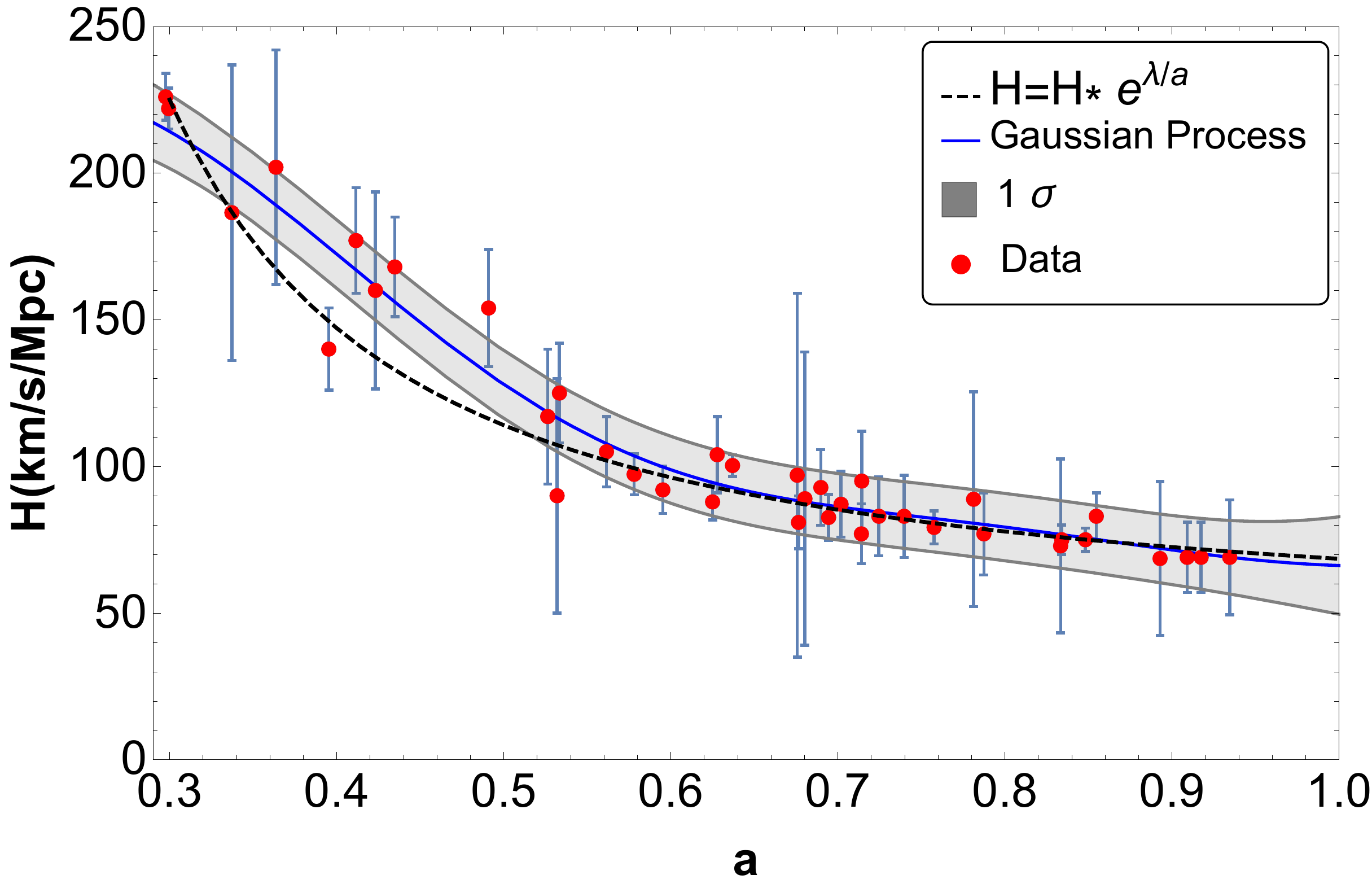}
    \caption{Same as Fig. \ref{fig:gaussiang1} but including the curve
    (dashed line) corresponding to parametrization 1 (Eq. (\ref{eq:Hmodel1})
    with $H_{*}=41.15$ km/s/Mpc and $\lambda= 0.51$).}
    \label{fig:Hmodel1}
\end{figure}

\begin{figure}
  \centering
  \includegraphics[width=0.93\linewidth]{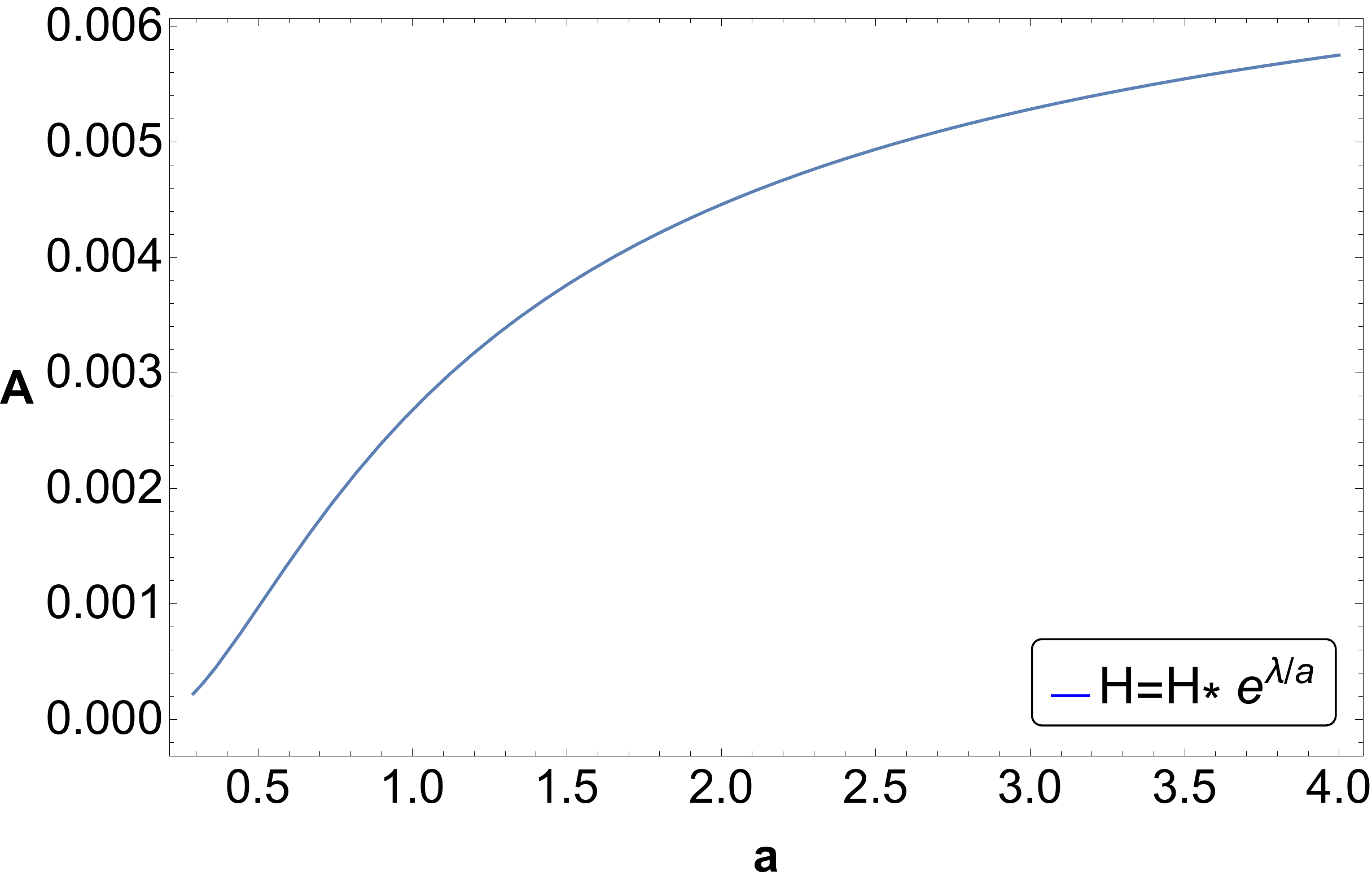}
  \caption{Area of the apparent horizon of parametrization 1 assuming $k = 0$. The graphs
  for  $k = +1$ and $k = -1$ are practically undistinguishable
  from this one and, therefore, not shown.}
  \label{fig:A&qmodel1}
\end{figure}

\noindent Figure \ref{fig:Hmodel1} contrasts Eq.
(\ref{eq:Hmodel1}), black dashed line, with the best fit to the
Hubble data. The dependence of the area of the apparent horizon on
the scale factor for $k =0$ is depicted in Fig.
\ref{fig:A&qmodel1}. The graphs for $k = +1$ and $k= -1$ show no
significant difference with this one; all three are practically
coincident (this is also true for parametrizations  2 and 3
considered below). From this graph we learn that ${\cal A}' \geq
0$ in the range of scale factor  considered and that ${\cal A}''
\leq 0$ from $a \simeq 0.5$ onward.
\subsection{Parametrization 2}
\begin{figure}
    \centering
    \includegraphics[width=0.45\textwidth]{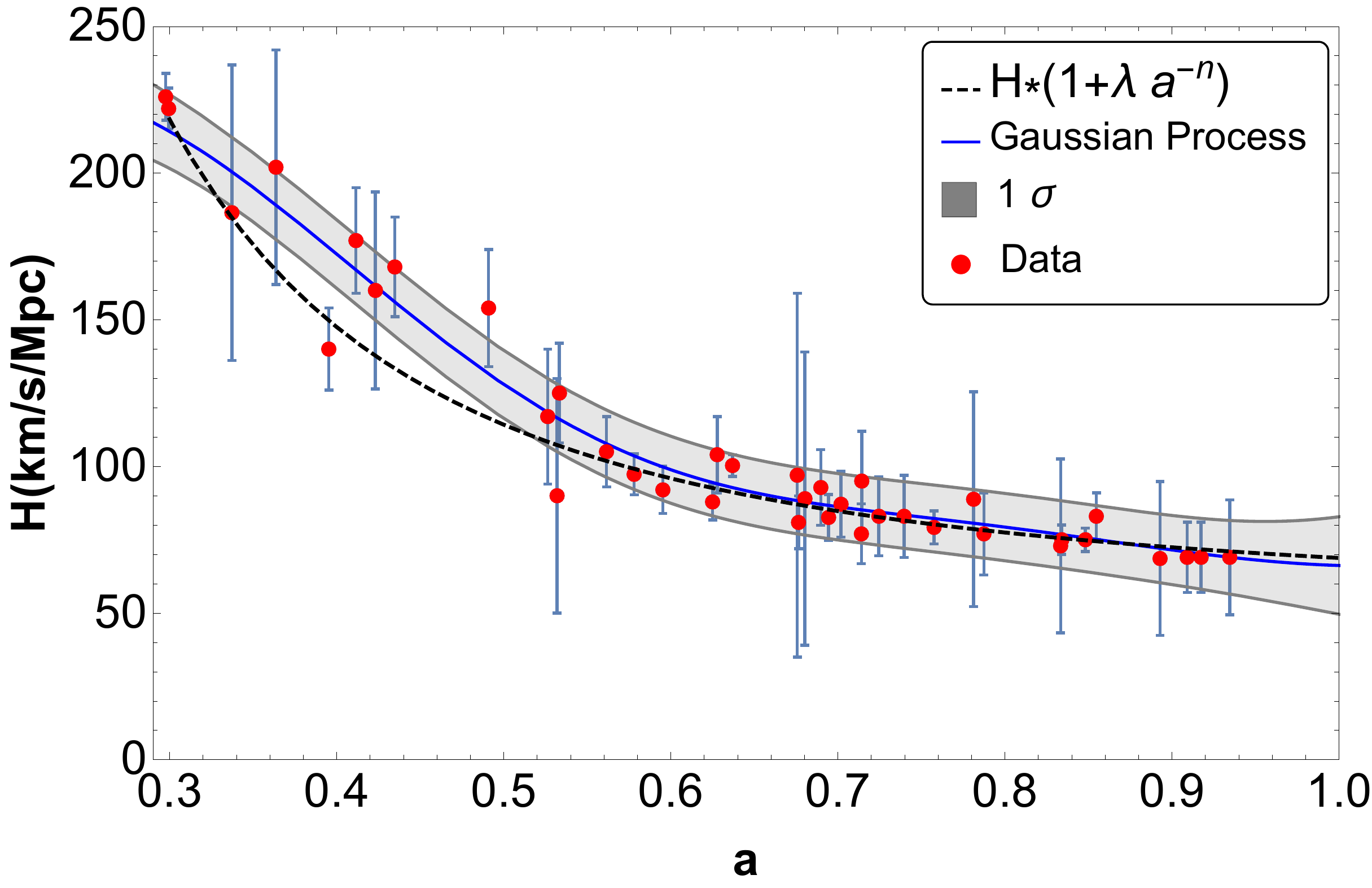}
    \caption{Same as Fig. \ref{fig:gaussiang1} but including  the curve
    (dashed line) of parametrization 2 (Eq. (\ref{eq:Hmodel2}) with
    $H_{*}=52.77$ km/s/Mpc, $\lambda= 0.30$ and $\, n=1.94$).}
    \label{fig:Hmodel2}
\end{figure}

\begin{figure}
\centering
  \includegraphics[width=0.45\textwidth]{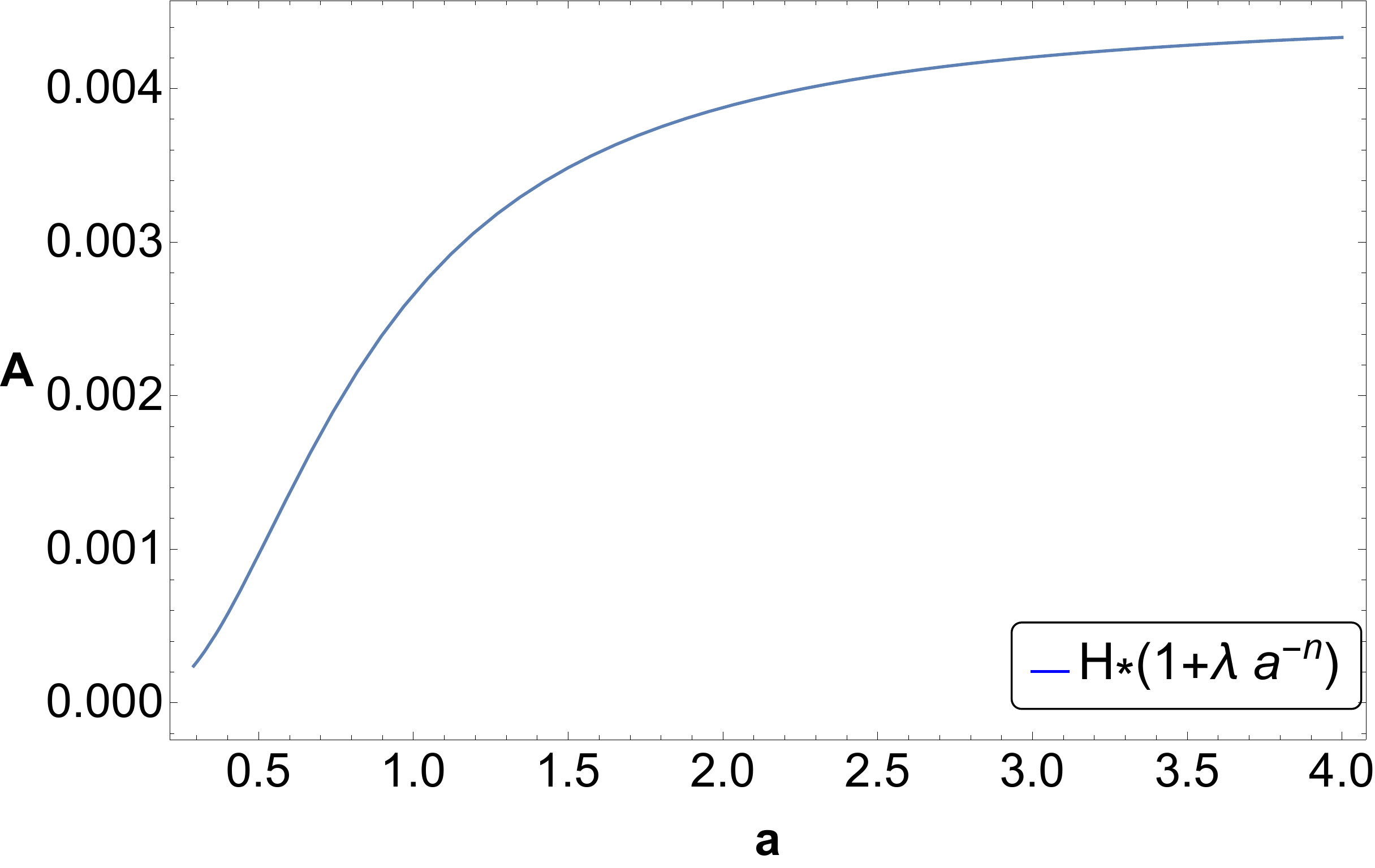}
  \caption{Area of the apparent horizon corresponding to parametrization 2.}
    \label{fig:A&qmodel2}
\end{figure}

\noindent The second parametrization is
\begin{equation}
H(a)=H_{*} (1+\lambda a^{-n})\, . \label{eq:Hmodel2}
\end{equation}
Proceeding as in the previous case, the best fit for the free
parameters  are found to be: $H_{*}=52.77 \pm 3.18$ km/s/Mpc,
$\lambda= 0.30 \pm 0.55$ and $n=1.94 \pm 0.09$. Thus, $t_0=13.9$
Gyr, $H_0=68.7$ km/s/Mpc, $q_{0}=-0.54$ and $a_{tr} = 0.53$.
Figure  \ref{fig:Hmodel2} compares this parametrization (black
dashed line) with the best fit to the Hubble data. The area
between the curves is $5.12$. Figure \ref{fig:A&qmodel2} shows the
evolution of the area of the apparent horizon.


\subsection{Parametrization  3}
\begin{figure}
    \centering
    \includegraphics[width=0.45\textwidth]{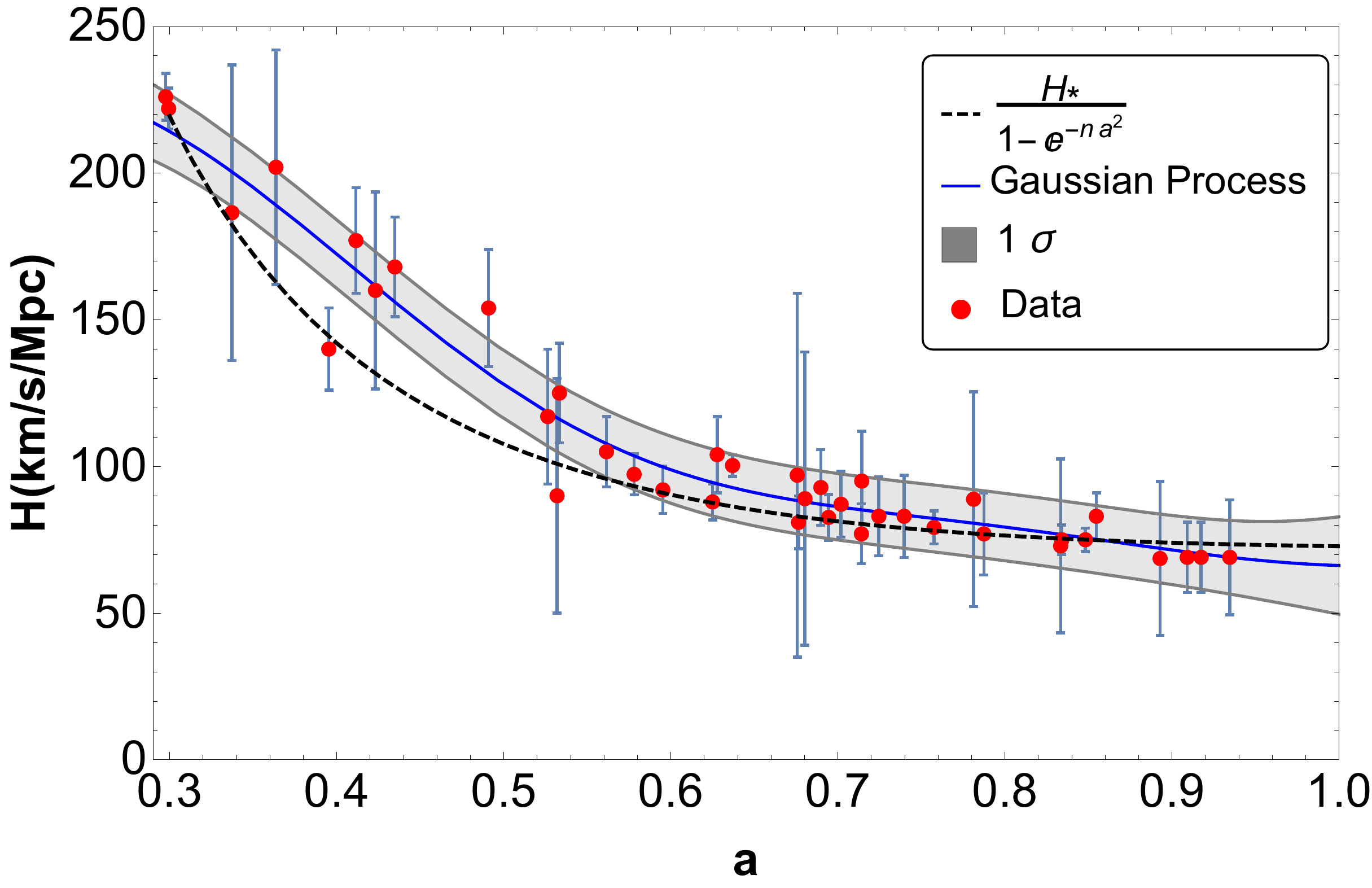}
    \caption{Same as Fig. \ref{fig:gaussiang1} but including the curve
    (dashed line) of parametrization 3 (Eq. (\ref{eq:Hmodel3}) with
    $H_{*}= 67.81$ km/s/Mpc and $n= 3.84$).}
    \label{fig:Hmodel3}
\end{figure}

\begin{figure}
\centering
  \includegraphics[width=0.93\linewidth]{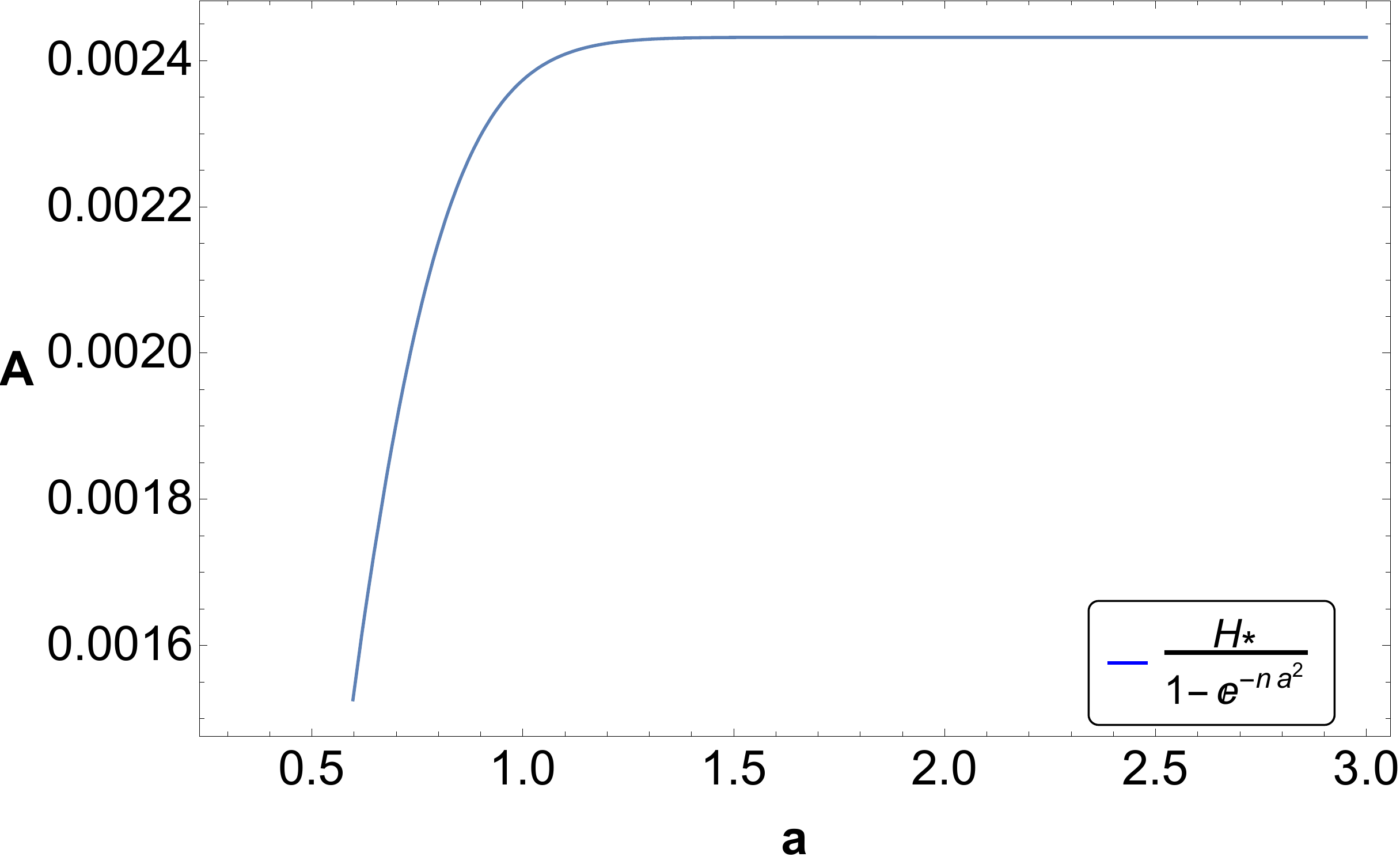}
  \caption{Area of the apparent horizon for parametrization 3.}
  \label{fig:A&qmodel3}
\end{figure}

\noindent As third parametrization, we propose
\begin{equation}
H(a)=\dfrac{H_{*}}{1\, -\, e^{-n a^{2}}}\, , \label{eq:Hmodel3}
\end{equation}
Proceeding as before yields, $H_{*}=67.81 \pm 3.58$ km/s/Mpc and
$n=3.84 \pm 0.32$. Thus, $t_{0}=14$ Gyr, $H_{0}=72.7$ km/s/Mpc,
$q_{0}=-0.89$, and $\, a_{tr} = 0.54$. Figure \ref{fig:Hmodel3}
contrasts the graph (black dashed line) corresponding to Eq.
(\ref{eq:Hmodel3}) with the best fit to the Hubble data. The area
between the curves is $7.79$. Figure  \ref{fig:A&qmodel3} shows
the evolution of the area of the horizon.
\[\]

\noindent The value for $q_{0}$ derived from these three essay
Hubble functions is quite consistent with the one measured by \cite{daly2008} ($q_{0} = -0.48 \pm 0.11$) by a model independent analysis.   These authors based their study on the
distances and redshifts to 192 supernovae, 30 radiogalaxies and 38
galaxy clusters. However, the quantity $a_{tr}$ that follow from
the said Hubble functions results slightly lower than the obtained
by Daly {\it et al.}, namely $a_{tr}= 0.56^{+0.10}_{-0.03}$.
\\  \

\noindent  All three  parametrizations share the features that,
regardless the sign of the curvature index $k$, the area, ${\cal
A}$, of the apparent horizon never decreases and that ${\cal A}''
\leq 0$ from some value of the scale factor on. This hints that
the universe described by each of them fulfills the second law of
thermodynamics and tend asymptotically to state of maximum entropy
of the order of $H^{-2}_{*}$ in Planck units.
\\  \

\noindent Nevertheless, Hubble
functions that comply with $H' < 0$ and $H'' > 0$ but fail to lead
to $\, {\cal A}'' \leq 0$, at late times (at least when $k = 0$)
can be proposed. Consider for instance  $\, H = H_{*} \, [\exp (\lambda/a) \, - \,
1]$ and $\, H = H_{*} \exp (-\lambda a)$ where $\lambda$ is a
positive definite constant. The entropy of
spatially-flat cosmological models whose expansion were governed
by any of these two functions would increase unbounded to
never reach a state of thermal equilibrium. However, as
previously remarked \citep{grg2}, both Hubble functions are at stark variance with
observation. The first one corresponds to a universe that never
accelerates. The second one to a universe that accelerates at
early times (i.e., for $a < \lambda^{-1}$) and decelerates forever
afterward.
\\   \

\noindent At this point one may wonder whether the evolution of the apparent horizon dictated by
the three Hubble's essay functions of above are qualitatively consistent with the corresponding evolution
predicted by those cosmological models that fit reasonably well the observational data.
This will be the subject of the next section.  As it will turn out, the answer to this question is
in the affirmative.

\section{Cosmological models dominated pressureless matter and dark energy}
\noindent In this section we briefly consider the evolution of the
area of the apparent horizon in a homogeneous and isotropic
universe governed by Einstein gravity and dominated by
pressureless matter (baryonic plus dark) and dark energy. We also
allow for the presence of a small spatial curvature. We first
study the case in which dark energy is in the form of a positive
cosmological constant, $\Lambda$, and then when it is given by a
scalar field with constant equation of state parameter $\, w_{X}
\neq -1$. In both cases we assume, in agreement with dynamical
measurements (see e.g., \cite{wendy2000} for a short review and \S
V in \cite{bartelmann}), that the matter component contributes to
the total energy budget by about 30 per cent. The rest is  in the
form of dark energy (either a cosmological constant or a scalar
field) plus spatial curvature. As stressed in \cite{ryan2018,
ooba2017, park2018a, park2018b} the latter may well be
non-negligible.
\\   \

\noindent Nowadays there is a discrepancy between local and
global measurements of the Hubble constant, $H_{0}$. The former
yield $73.24 \pm 1.74$ km/s/Mpc \citep{Riess2016} while the latter, based on the spatially flat
$\Lambda$CDM model, give $\, 67.8 \pm 0.9$ km/s/Mpc
\citep{planck2015}. Below, we will use in turn the result of Riess  as
well as the prior $68 \pm 2.8$ km/s/Mpc employed by \cite{ryan2018}. As we will see, no major difference arise with regard to the overall evolution of the area of the horizon.
At any rate, both values are somewhat above the best fit obtained
by the Gaussian Process (blue solid  line in Fig.
\ref{fig:gaussiang1}). As input data we will employ the cosmological parameter
values obtained in \cite{ryan2018} by constraining several
simple cosmological models based in Einstein gravity (both with
and without spatial curvature) using 31 $\, H(z)$ data in the
redshift range $\, 0.07 \leq z \leq 1.965$ and 11 baryon acoustic
oscillation distance measurements.

\subsection{$\Lambda$CDM models}
\noindent For this kind of models Friedamann's equation reads
\begin{equation}
H(a)=H_0 \sqrt{\Omega_{m0}\, a^{-3}\, + \, \Omega_{k0}\, a^{-2}\,
+ \, \Omega_{\Lambda}},
\label{eq:friedmann-LCDM1}
\end{equation}
where $\Omega_{m0} $ and $\Omega_{k0} = 1-
\Omega_{m0}-\Omega_{\Lambda}$ denote the present value of the
density parameters of matter and spatial curvature, respectively,
and $\Omega_{\Lambda}$ the parameter density associated to the
cosmological constant. From (\ref{eq:friedmann-LCDM1}) it is seen
that at large times the area of the apparent horizon is of the
order of $\, H^{-2}_{0} \Omega^{-1}_{\Lambda}$.
\\  \

\begin{figure}
    \centering
    \includegraphics[width=0.45\textwidth]{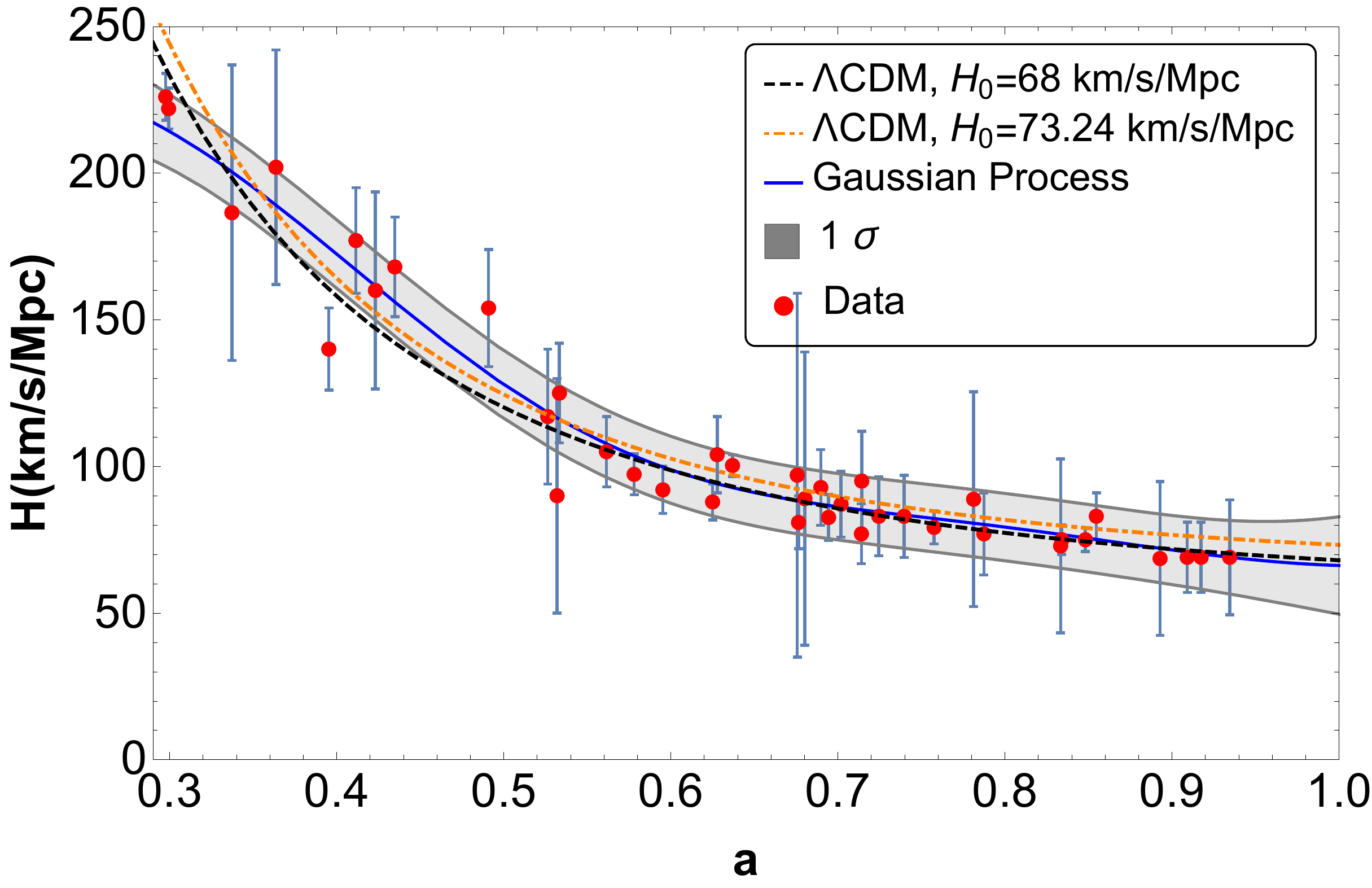}
    \caption{The Hubble factor for the two $\Lambda$CDM models contrasted with the best fit
    to the Hubble data in  table \ref{table:H(z)data} using the Gaussian Process.}
\label{fig:gaussian&LCDM1}
\end{figure}

\noindent We study two $\Lambda$CDM cases:
\begin{itemize}
\item First case, $H_0= 68 \pm 2.8$ km/s/Mpc,
$\Omega_{m0}=0.29$
 and $\Omega_{\Lambda0}=0.68$ (see  black dashed line in Fig. \ref{fig:gaussian&LCDM1}). Thus, $t_0=13.9$
Gyr, $q_0=-0.54$ and $a_{tr} = 0.60$. The area between the curves is $3.25$.
\item Second case, $H_0=73.24 \pm 1.74$ km/s/Mpc, $\Omega_{m0}=0.30$ and
$\Omega_{k0}=-0.07$ (see orange dot-dashed  line in Fig. \ref{fig:gaussian&LCDM1}). Thus,
$t_0=13.2$ Gyr, $q_0 = -0.62$ and  $\, a_{tr} = 0.58$. The
area between the curves is $3.86$.
\end{itemize}

\begin{figure}
\centering
  \includegraphics[width=0.93\linewidth]{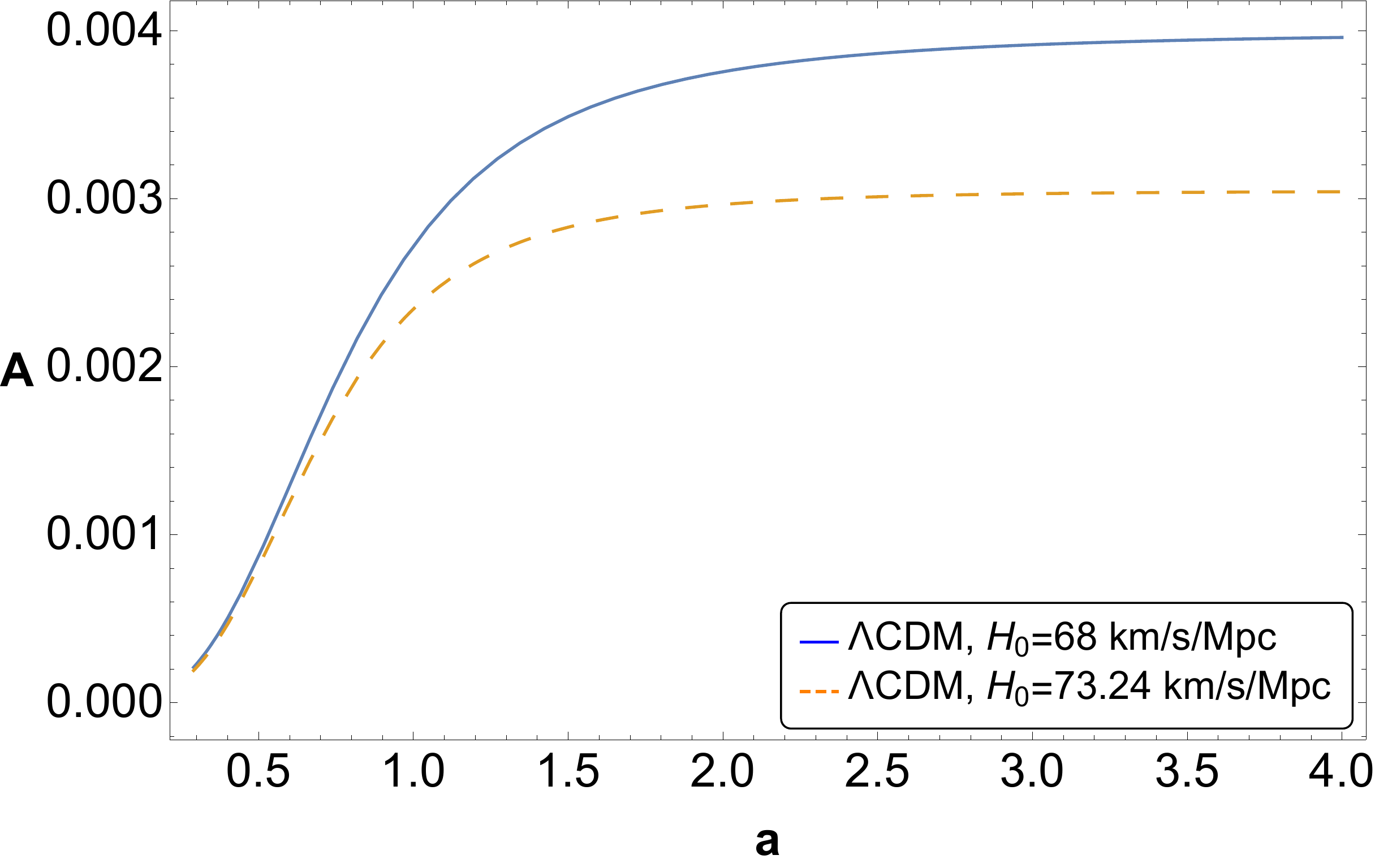}
  \caption{Area of the apparent horizon in the $\Lambda$CDM models.}
    \label{fig:A&qLCDM}
\end{figure}

\noindent The evolution of the area of the apparent horizon is shown, in each case, in
Fig. \ref{fig:A&qLCDM}.

\subsection{flat-XCDM models}
\begin{figure}
    \centering
    \includegraphics[width=0.45\textwidth]{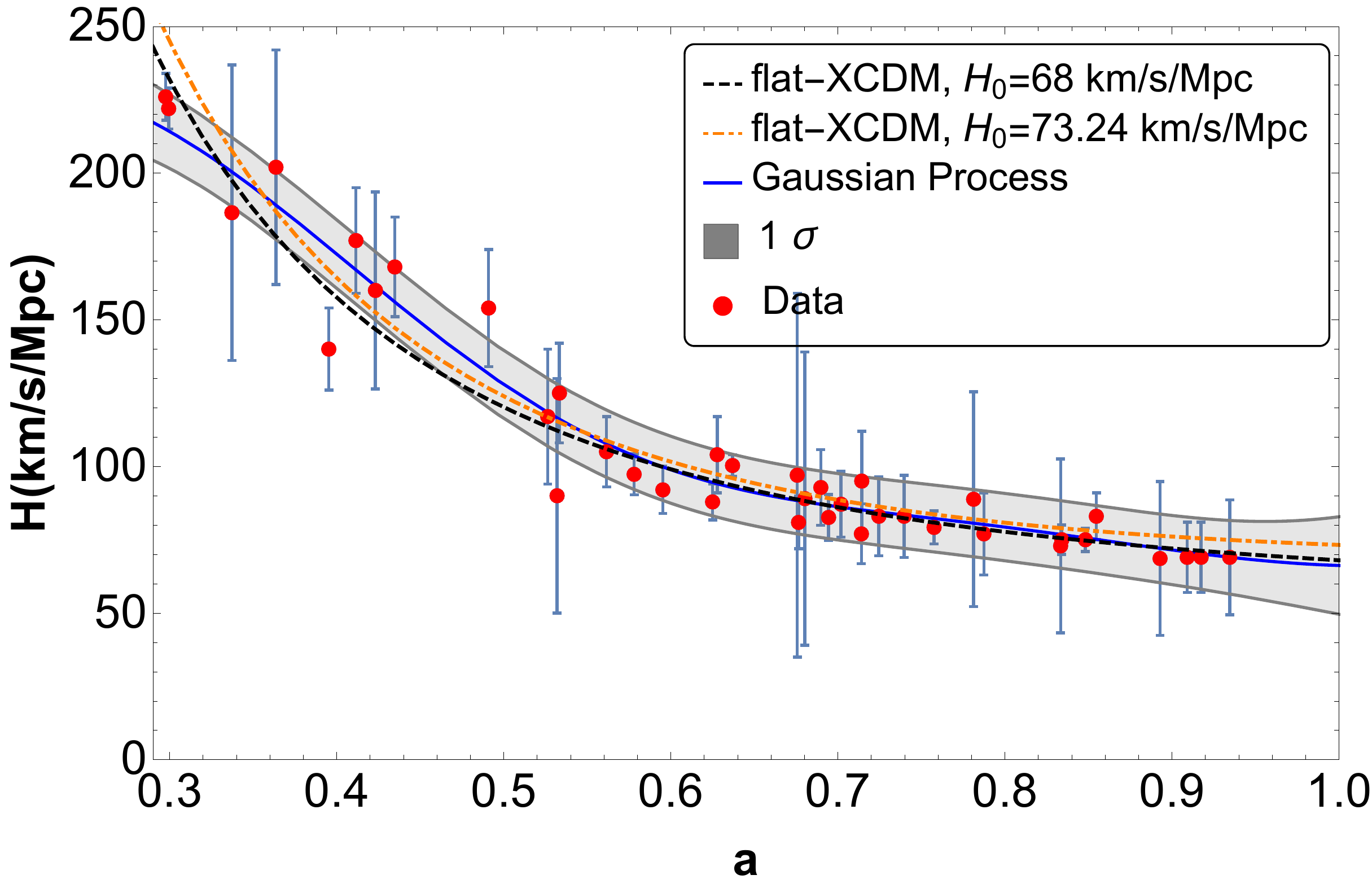}
    \caption{Same as Fig. \ref{fig:gaussian&LCDM1} but for the two flat XCDM models.}
    \label{fig:gaussian-flatXCDM}
\end{figure}

\begin{figure}
\centering
  \includegraphics[width=0.93\linewidth]{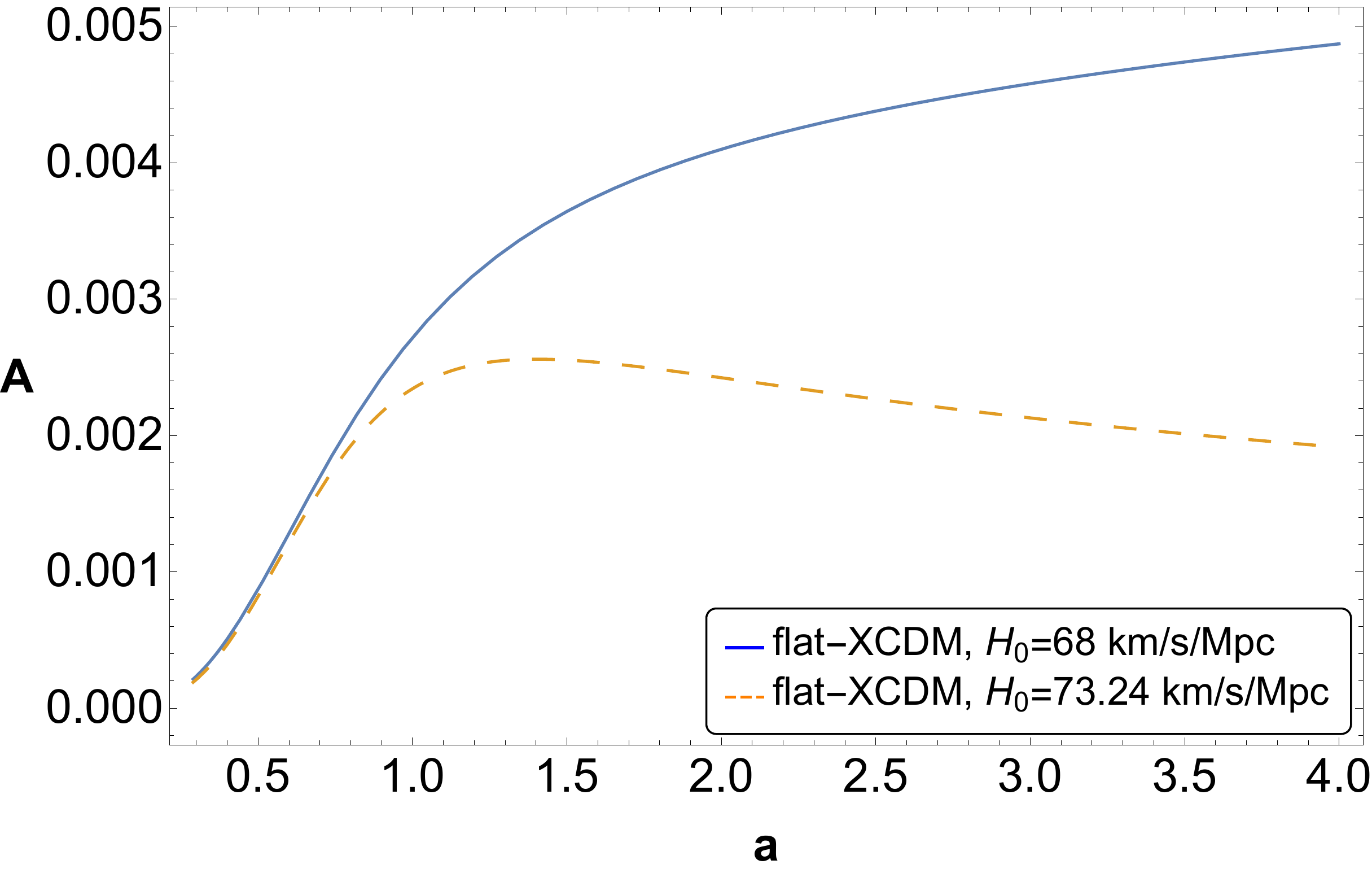}
  \caption{Area of the apparent horizon for the flat XCDM models.}
  \label{fig:A-flatXCDM}
\end{figure}

\noindent We consider a generic dark energy model that, from the
phenomenological viewpoint, essentially differs at first order
from the $\Lambda$CDM in that the equation of state parameter,
$w_{X} \equiv p_{X}/\rho_{X}$, is a constant  different from $-1$.
We study in turn two flat-XCDM cases where
\begin{equation}
H(a) = H_{0} \, \sqrt{\Omega_{m0}\, a^{-3} \, + \, \Omega_{X0}\,
a^{-3(1+w_X)}} \quad \quad (\Omega_{X0} = 1 \, - \, \Omega_{m0}).
\label{eq:friedmann-XCDM1}
\end{equation}
\begin{itemize}
\item First case, $H_0= 68 \pm 2.8 \pm 0.9$ km/s/Mpc,
$\Omega_{m0}=0.29$ and $w_X=-0.94$ (see black dashed  line in Fig.
\ref{fig:gaussian-flatXCDM}). Thus, $t_0=13.9$ Gyr, $q_0=-0.50$
and $a_{tr} = 0.59$.  The area between the curves is $3.11$.
\item Second case, $H_0=73.24 \pm 1.74$ km/s/Mpc,
$\Omega_{m0}=0.29$  and $w_X=-1.13$ (see orange dot-dashed  line
in Fig. \ref{fig:gaussian-flatXCDM}). Thus, $t_0=13.3$ Gyr,
$q_0=-0.70$  and $\, a_{tr} = 0.60$.  The area between the curves
is $3.57$.  Note  \textemdash see Fig. \ref{fig:A-flatXCDM}
\textemdash that the area of the apparent horizon begins
increasing but at some point (as soon as the dark energy  takes
over) it decreases. This violation of the second law (which occurs
because, in this case, the dark energy is of ``phantom" type,
$w_{X} < -1$) goes hand in hand with the fact that phantom fields
present classical \citep{dabrowski} and quantum instabilities
\citep{q-instabilities1, q-instabilities2} that render them
implausible.
\end{itemize}

\noindent In both cases the evolution of the area of the horizon is shown
Fig. \ref{fig:A-flatXCDM}.
\subsection{Non-flat XCDM models}
\begin{figure}
    \centering
    \includegraphics[width=0.45\textwidth]{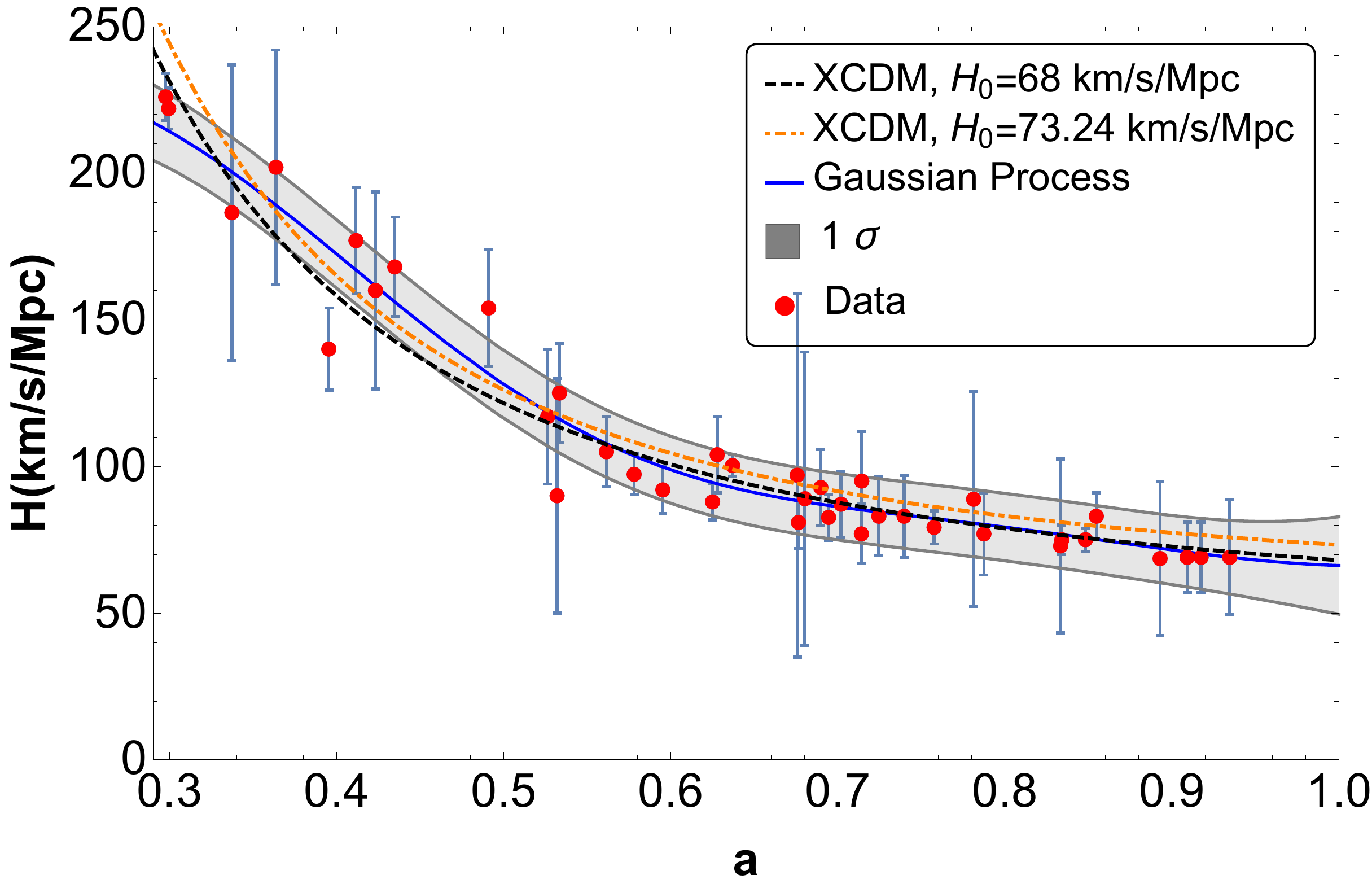}
    \caption{Same as Fig. \ref{fig:gaussian&LCDM1} but for the non-flat the XCDM models.}
    \label{fig:gaussian-nfXCDM}
\end{figure}

The Hubble factor of spatially curved XCDM models can be written as
\begin{equation}
H(a)=H_0 \, \sqrt{\Omega_{m0} \, a^{-3}\, + \, \Omega_{k0} \,
a^{-2} \, + \, \Omega_{X0} \, a^{-3(1+w_{X})}} \quad \, ,
\label{eq:friedmann-XCDM2}
\end{equation}
\begin{equation*}
(\Omega_{X0} = 1 \, -\, \Omega_{m0} \, - \, \Omega_{k0}).
\end{equation*}
\begin{itemize}
\item First case, $\, H_0= 68 \pm 2.8 $ km/s/Mpc,
$\Omega_{m0}=0.31$, $\Omega_{k0}=-0.18$ and $w_{X}=-0.76$
(see black dashed  line in Fig. \ref{fig:gaussian-nfXCDM}). Thus,
$t_0=13.8$ Gyr, $q_0=-0.40$    and $a_{tr} = 0.58$. The area
between the curves is $2.99$.  \item Second case, $H_0=73.24 \pm 1.74$
km/s/Mpc, $\Omega_{m0}=0.32$, $\Omega_{k0}=-0.21$, $w_{X}=-0.84$
(see orange dot-dashed  line in Fig. \ref{fig:gaussian-nfXCDM}).
Thus, $t_0=13$ Gyr, $q_0=-0.52$ and $\, a_{tr} = 0.57$. The
area between the curves is $\, 4.29$.
\end{itemize}

\begin{figure}
\centering
  \includegraphics[width=0.93\linewidth]{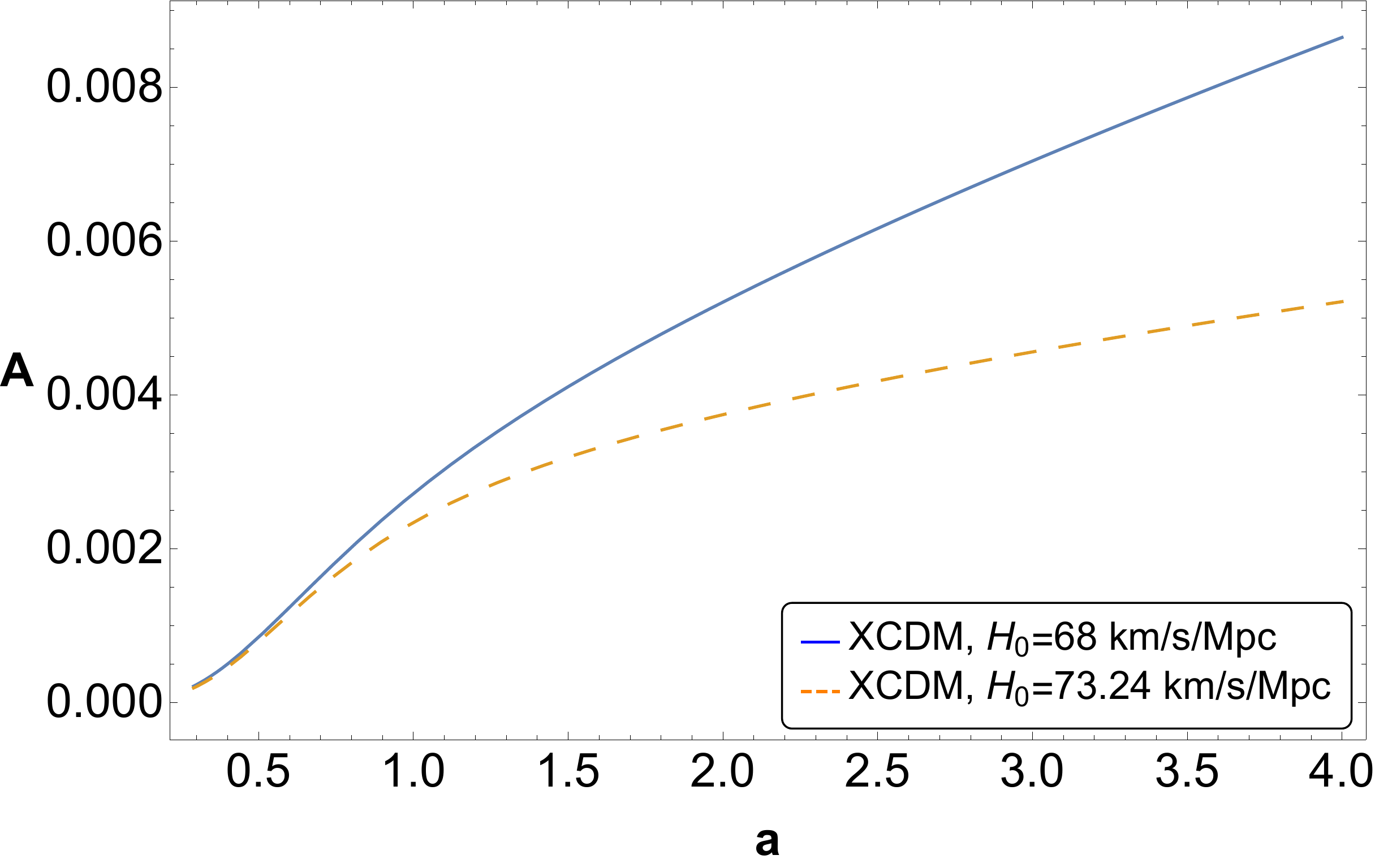}
  \caption{Evolution of the area of the horizon the non-flat the XCDM model.}
    \label{fig:A-nfXCDM}
\end{figure}

\noindent The evolution of the area of the apparent horizons is depicted in Fig. \ref{fig:A-nfXCDM}.
\\   \

\noindent Notice that the cosmological models considered in this
section fit better (by about a factor of 2) the best fit (blue
solid line) to the Hubble's data set in Fig. \ref{fig:gaussiang1}
than any of the three parametrizations of section 3 (i.e., the
former have a lower value for the area between the curves than the
latter) even though no fitting process to the said data has been
applied to the cosmological models. This was to be expected with
regard to the Hubble essay functions 1 and 3 because the  models
of this section have one or two  more free parameters.
\\   \

\noindent We have just considered a handful of simple cosmological models that fit observation reasonably well
 \textemdash see Ref. \cite{ryan2018}. More sophisticated  models, as those beyond Einstein gravity, also deserve
consideration. However, generally speaking, in that case the fit of the model parameters to the
observational data is significantly affected by systematics.
\\  \

\noindent We conclude this section by noting that the apparent horizon of homogeneous and isotropic
cosmological models, regardless of their spatial curvature,  that fit current observation and are classical
and quantum-mechanically stable comply with the same  conditions (${\cal A'} > 0$ and ${\cal A''} \leq 0$,
the latter at least from some value of the scale factor onward) than the three essay
Hubble functions (Eqs. (\ref{eq:Hmodel1}), (\ref{eq:Hmodel2}) and (\ref{eq:Hmodel3})) of
section III. These two conditions ensure that the corresponding cosmological model respects
the second law. This suggests that from this preliminary $H(z)$ dataset, however scarce and of
limited  quality, one may glimpse that the universe fulfills the said law. Therefore,
from this point of view, it appears to behave as a normal thermodynamic system. Recently, this
conclusion was also attained by an altogether different route, namely by the analysis of
the statistical fluctuations of the flux of energy on the apparent horizon
\citep{jp-diego2018}. As it turns out, the strength of the fluctuations decrease
with the area of the horizon and increase with the temperature of the latter.
Just as in systems in which gravity is absent \citep{Landau-Lifshitz}.

\begin{table}
\caption{Hubble's parameter vs. redshift \& scale factor.}
\label{table:H(z)data}
\renewcommand{\tabcolsep}{0.7pc} 
\renewcommand{\arraystretch}{0.7} 
\begin{tabular}{@{}lllll}
\hline \hline
  $\;\; z$ & $\;\; a$    &  $ H(z) \;$ (km s$^{-1}$ Mpc$^{-1}$) &  Ref. \\
\hline
$0.07$      & $0.93$       & $ \; \qquad 69     \pm 19.6 $      &   \cite{zhang2014} \\
$0.09$      & $0.92$       & $ \; \qquad 69     \pm 12 $      & \cite{simon2005} \\
$0.100$     & $0.91$       & $ \; \qquad 69     \pm 12 $      & \cite{simon2005} \\
$0.120$     & $0.89$       & $ \; \qquad 68.6     \pm 26.2$       & \cite{zhang2014} \\
$0.170$     & $0.85$       & $ \; \qquad 83     \pm 8$       & \cite{simon2005} \\
$0.179$     & $0.84$       & $ \; \qquad 75     \pm 4$       & \cite{moresco2012} \\
$0.199$     & $0.83$       & $ \; \qquad 75     \pm 5$        & \cite{moresco2012} \\
$0.200$     & $0.83$       & $ \; \qquad 72.9     \pm 29.6$        &  \cite{zhang2014} \\
$0.270$     & $0.79$       & $ \; \qquad 77     \pm 14$      & \cite{simon2005} \\
$0.280$     & $0.78$       & $ \; \qquad 88.8     \pm 36.6$      & \cite{zhang2014} \\
$0.320$     & $0.75$       & $ \; \qquad 79.2   \pm 5.6$     & \cite{cuesta2016}\\
$0.352$     & $0.74$       & $ \; \qquad 83     \pm 14$      & \cite{moresco2012} \\
$0.3802$    & $0.72$       & $ \; \qquad 83     \pm 13.5$      & \cite{moresco2012} \\
$0.400$     & $0.71$       & $ \; \qquad 95     \pm 17$      & \cite{simon2005} \\
$0.4004$    & $0.71$       & $ \; \qquad 77     \pm 10.2$      & \cite{moresco2012} \\
$0.4247$    & $0.70$       & $ \; \qquad 87.1     \pm 11.2$      & \cite{moresco2012} \\
$0.440$     & $0.69$       & $ \; \qquad 82.6   \pm 7.8$     & \cite{blake2012} \\
$0.4497$    & $0.69$       & $ \; \qquad 92.8   \pm 12.9$     & \cite{moresco2012} \\
$0.470$     & $0.68$       & $ \; \qquad 89   \pm 50$     & \cite{ratsim} \\
$0.4783$    & $0.68$       & $ \; \qquad 80.9   \pm 9$     & \cite{moresco2012} \\
$0.480$     & $0.68$       & $ \; \qquad 97     \pm 62$      & \cite{stern2010} \\
$0.570$     & $0.64$       & $ \; \qquad 100.3  \pm 3.7$     & \cite{cuesta2016} \\
$0.593$     & $0.63$       & $ \; \qquad  104   \pm 13$      & \cite{moresco2012} \\
$0.600$     & $0.63$       & $ \; \qquad 87.9   \pm 6.1$     & \cite{blake2012} \\
$0.680$     & $0.60$       & $ \; \qquad 92     \pm 8$       & \cite{moresco2012} \\
$0.730$     & $0.58$       & $ \; \qquad 97.3   \pm 7 $      & \cite{blake2012} \\
$0.781$     & $0.56$       & $ \; \qquad 105    \pm 12$      & \cite{moresco2012} \\
$0.875$     & $0.53$       & $ \; \qquad 125    \pm 17$      & \cite{moresco2012} \\
$0.880$     & $0.53$       & $ \; \qquad 90     \pm 40$      & \cite{stern2010} \\
$0.900$     & $0.52$       & $ \; \qquad 117    \pm 23$      & \cite{simon2005} \\
$1.037$     & $0.49$       & $ \; \qquad 154    \pm 20 $     & \cite{moresco2012} \\
$1.300$     & $0.43$       & $ \; \qquad 168    \pm 17 $     & \cite{simon2005} \\
$1.363$     & $0.42$       & $ \; \qquad 160    \pm 33.6$    & \cite{moresco2015}\\
$1.430$     & $0.41$       & $ \; \qquad 177    \pm 18$      & \cite{simon2005}\\
$1.530$     & $0.40$       & $ \; \qquad 140    \pm 14$      & \cite{simon2005}\\
$1.750$     & $0.36$       & $ \; \qquad 202    \pm 40$      & \cite{simon2005}\\
$1.965$     & $0.34$       & $ \; \qquad 186.5  \pm 50.4$    & \cite{moresco2015}\\
$2.340$     & $0.30$       & $ \; \qquad 222    \pm 7 $      & \cite{delubac2014}\\
$2.360$     & $0.30$       & $ \; \qquad 226    \pm  8$      & \cite{font-ribera2014}\\
\hline
\end{tabular}\\
 \end{table}

\section{Concluding remarks}

\noindent The recent years have witnessed a reinforcement of the
connection between gravity and thermodynamics. As two salient
landmarks we mention the discovery that fully gravitationally
collapsed objects possess a well defined temperature and entropy
\citep{hawking-1975,hawking-1976}, and the realization that the gravity field
equations can be viewed as thermodynamic equations of state
\citep{jacobson-1995,paddy-2005}. In this context it is natural to
ask whether the universe behaves as a normal thermodynamic system;
that is to say, whether it satisfies the laws of thermodynamics;
most importantly the second law. Actually, this amount to consider
whether the aforesaid law also holds at cosmic scales.
\\   \

\noindent To answer this question we assumed the universe
homogeneous and isotropic at sufficiently large scales and,  based
on the history of the Hubble function as specified in table
\ref{table:H(z)data}, made a kinematic analysis independent of any
cosmological model. In section II, by applying the Gaussian
Process we obtained a best fit to the data (blue solid line in
Fig. \ref{fig:gaussiang1}). The latter suggests that $H'(a) < 0$
and $H'' (a) > 0$. Then, in section III  we essayed three simple
Hubble functions (Eqs. (\ref{eq:Hmodel1}), (\ref{eq:Hmodel2}) and
(\ref{eq:Hmodel3})) that comply with these two inequalities, and
are consistent with the second law as applied to the area of the
apparent horizon (i.e., ${\cal A'} > 0 \, $ and $\, {\cal A''} <
0$), and reproduce reasonably well  (especially for $a \geq 0.55$)
the best fit to the Hubble data \textemdash Figs.
\ref{fig:Hmodel1}, \ref{fig:Hmodel2} and \ref{fig:Hmodel3}. As
illustrated by two examples, Hubble functions that imply  $\,
{\cal A''} > 0$ in the long run strongly disagree with observation
at the background level. In section IV we contrasted the evolution
of the Hubble function  predicted by some simple, flat as well as
non-flat, cosmological models that fit observation reasonably well
\citep{ryan2018}  with the best fit to the Hubble dataset (Figs.
\ref{fig:gaussian&LCDM1}, \ref{fig:gaussian-flatXCDM}  and
\ref{fig:gaussian-nfXCDM}). These fits are only somewhat better
than those of the three parametrizations  proposed in section III.
Except for the second flat-XCDM model, that is of ``phantom" type
and presents instabilities at the classical and quantum level, the
area of the apparent horizon is always increasing and with second
derivative negative in the long run. In this regard, it mimics the
evolution of the area of the apparent horizon associated to the
three essay Hubble functions of section III.
\\   \

\noindent Our overall conclusion is that the second law of
thermodynamics seems to be obeyed by the universe we observe today
and, therefore, that this law also holds at cosmic scales.
However, this conclusion, extracted from a scarce number of Hubble
data of not great quality, cannot be but preliminary.
Nevertheless, we hope that in the not distant future a much more
ample set of data of better quality will be available whence we
will make able to reach a firmer conclusion. This improvement in
the quality and number of data may arise not only from a
refinement of the current techniques but also from measuring the
drift of the redshift of distant sources
\citep{sandage-1962, loeb-1998}. This seems feasible thanks to the
future European Extreme Large Telescope alongside advanced
spectographs like \cite{codex}.
\\   \

\noindent Finally, as is apparent, our approach is purely classical in nature. Nevertheless, 
it is worthy to emphasize that it is also robust against quantum fluctuations since, 
as demonstrated by \cite{oshita2018generalized}, these do not invalidate the generalized 
second law because cosmological decoherence prevents it in the case of 
de Sitter expansion.


\section*{Acknowledgements}
M. G.-E. acknowledges support from a PUCV doctoral
scholarship and DI-VRIEA-PUCV for financial support.




\bibliographystyle{mnras}
\bibliography{bio} 

\bsp    
\label{lastpage}
\end{document}